\documentclass[a4paper,english,aps,prl,twocolumn,floatfix,showpacs,amsfonts,amssymb,superscriptaddress]{revtex4} 
\usepackage[T1]{fontenc}
\usepackage[latin1]{inputenc}
\usepackage{graphics}
\usepackage{amssymb}
\usepackage{amsmath}
\usepackage{overpic}

\newcommand{\be}{\begin{equation}}
\newcommand{\ee}{\end{equation}}
\newcommand{\bea}{\begin{eqnarray}}
\newcommand{\eea}{\end{eqnarray}}

\def\l{\lambda}
\def\th{\theta}

\def\Mt{\tilde M}

\def\D{\Delta}
\def\O{\Omega}
\def\L{\Lambda}

\def\ra{\rightarrow}

\def\pd{\partial}
\def\nb{\nabla}
\def\bk{{\bf k}}
\def\bl{{\bf l}}

\def\br{{\bf r}}

\def\bA{{\bf A}}
\def\bB{{\bf B}}
\def\bH{{\bf H}}

\def\bJ{{\bf J}}
\def\bM{{\bf M}}

\def\DD{{\cal{D}}}
\def\GG{{\cal{G}}}

\def\nn{\nonumber}
\def\lb{\label}
\def\pref#1{(\ref{#1})}

\newcount\bozza \bozza=0
\ifnum\bozza=1
\newdimen\shift \shift=-2truecm
\def\lb#1{%
{\label{#1}\rlap{\kern\shift{$\scriptstyle#1$}}}}
\else\def\lb#1{\label{#1}} \fi

\begin{document}

\title{Sine-Gordon description of Beresinskii-Kosterlitz-Thouless
physics\\ at finite magnetic field}

\author{L.~Benfatto}
\affiliation
{Centro Studi e Ricerche ``Enrico Fermi'', via Panisperna 89/A, I-00184,
  Rome, Italy} 
\affiliation
{CNR-SMC-INFM and Department of Physics, University of Rome ``La
  Sapienza'',\\ Piazzale Aldo Moro 5, I-00185, Rome, Italy}

\author{C.~Castellani}
\affiliation
{CNR-SMC-INFM and Department of Physics, University of Rome ``La
  Sapienza'',\\ Piazzale Aldo Moro 5, I-00185, Rome, Italy}

\author{T.~Giamarchi}
\affiliation{DPMC- MaNEP University of Geneva, 24 Quai Ernest-Ansermet CH-1211
Gen\`eve 4, Switzerland}

\date{\today}

\begin{abstract}
The Beresinskii-Kosterlitz-Thouless (BKT) physics of vortices in
two-dimensional superconductors at finite magnetic field is investigated by
means of a field-theoretical approach based on the sine-Gordon model. This
description leads to a straightforward definition of the field-induced
magnetization and  shows that the persistence of non-linear effects at
low fields above the transition is a typical signature of the
fast divergence of the correlation length within the BKT theory.

\end{abstract}


\maketitle 

The Beresinskii-Kosterlitz-Thouless (BKT) transition
\cite{berezinsky_nolongrange_2d}, namely the possibility to have a phase
transition with a vanishing order parameter but algebraic decay of the
correlations, is undoubtedly one of the most fascinating aspects of
collective phenomena. It finds experimental realizations in a wide range of
systems, as superfluids or superconducting (SC)
films\cite{review_minnaghen,armitage_films,triscone_fieldeffect,lesueur_films}
and recently cold atomic systems \cite{hadzibabic_ENS_experiment_BKT}. One
of the key ingredients of the BKT transition is the existence of vortices,
that unbind in the high-temperature phase leading to an exponential decay
of the correlations. In order to treat such unbinding transition a very
fruitful analogy was to represent the vortices as charges performing a
Debye-Huckel screening transition in a neutral Coulomb-gas problem, for
which a renormalization group procedure can be implemented
\cite{review_minnaghen}.

One specially interesting extension of the BKT transition is when a
magnetic field is present, which will impose a population of vortices with
a given vorticity in the system.  This has found recent experimental
application to thin films\cite{armitage_films,lesueur_films} or layered
high-T$_c$ superconductors\cite{li_magn}.  Even in cold atomic systems, a
magnetic field can be mimicked by imposing a rotation on the condensate
\cite{hadzibabic_ENS_experiment_BKT,cazalilla_coupled_XY_cold}.  For all
these systems it is thus crucial to predict theoretically how the magnetic
field will affect the BKT transition and the various physical observables.

Due to the strong interest of such a question, this problem has been
addressed in the past
\cite{review_minnaghen,doniach_films,minnaghen_finitefield,sondhi_kt}.
Unfortunately, contrarily to the case of the $\bB=0$ transition, the
efforts have been partly unsatisfactory. In particular most of the
literature on the subject rested on extending the mapping to the
Coulomb-gas problem, where the effects of the magnetic field can be
incorporated as an excess of positive charges. However this mapping gives
the physical observables as a function of the magnetic
induction $\bB$ instead of the magnetic field $\bH$, which is not
convenient to describe the physics at low applied field. 

An alternative approach to the BKT transition, which is of course well
known for $\bB=0$, is to use the mapping onto the sine-Gordon
problem\cite{review_minnaghen,giamarchi_book_1d}, which
was reviewed recently both in the context of quasi-2D superconductors 
\cite{benfatto_BKT_in_cuprates,nandori_kt} and cold atomic systems
\cite{cazalilla_coupled_XY_cold}.  In this Letter we show that this
description 
provides a very simple and physically transparent way to deal with the
finite magnetic field case. In our scheme the physical observables have a
straightforward definition, and the role of both $\bB$ and $\bH$ is
clarified. In addition we also present a variational calculation of the
field-induced diamagnetism in thin films. It leads to a detailed
description of the Meissner phase below $T_{BKT}$ and of the appearance
above $T_{BKT}$ of a non-linear magnetization at relatively low fields, in
contrast to what expected from standard Ginzburg-Landau (GL) SC
fluctuations\cite{koshelev_gaussian}.

As a starting model we consider the $XY$ model for the phase of a 2D
superconductor\cite{review_minnaghen}
\be
\lb{ham}
H=J\sum_{<i,j>} [1-\cos(\th_i-\th_j-F_{ij})].
\ee
Here $\theta_{i,j}$ is the SC phase on two nearest-neighbor sites $(i,j)$
of a coarse-grained 2D lattice, $J={\Phi_0^2d}/16\pi^3\l^2$ is the 2D
superfluid stiffness for a film of thickness $d$ and in-plane penetration
depth $\l$, and we employed a minimal-coupling scheme for the vector
potential $\bA$, with $F_{ij}=(2\pi/\Phi_0)\int_i^j\bA\cdot d\bl$, and
$\Phi_0=hc/2e$ the flux quantum. Due to the periodicity of $H$ when
$\theta\ra \theta+2\pi$, beyond long-wavelength phase excitations where
$\theta_i-\theta_j\approx a\nb \theta$ varies smoothly on the lattice scale
$a$, vortex configurations are allowed where $\oint \nabla \theta =\pm
2\pi$ over a closed loop. They emerge clearly by performing the standard
dual mapping of the model \pref{ham}\cite{jose}. This allows us to write
the partition function of the system
as a functional integral over a scalar field $\phi$ as $Z=\int \DD\phi
e^{-S_B}$, 
\be
\lb{sb}
S_B\!=\negthickspace\!\int \!\!d\br dz\!\!
\left[\frac{(\nb \phi)^2}{2\pi K} \!-\!\frac{g}{\pi a^2}\cos 2\phi
+\frac{2 i}{\Phi_0}\bA\!\cdot\!(\nb \times   \hat z\phi)
\right]\!\delta(z),\,
\ee
where $\phi$ depends on the in-plane coordinates $\br$ only while $\bA$
depends in general also on the $z$ coordinate. The $\delta(z)$ function
gives the proper boundary conditions for a truly 2D case (where there is no
SC current outside the plane). In the physical case of a SC film of
thickness $d$ we assume that the sample quantities are averaged over
$|z|<d/2$. In Eq.\ \pref{sb} we defined $K=\pi J/k_BT$ and $g= 2\pi
e^{-\beta \mu}$, where $\mu$ is the chemical potential of the vortices and
$e^{-\beta\mu}$ their fugacity ($\beta=1/k_B T$).  While in the $XY$ model
$\mu/J$ is fixed, $\mu_{XY}\simeq \pi^2 J/2$, we consider it as an
independent variable\cite{benfatto_BKT_in_cuprates}.
In the dual representation \pref{sb} of the XY model
\pref{ham} the cosine term accounts for vortex excitations: indeed,
since $\phi$ is the dual field of $\theta$, a vortex, which is a $\pm2\pi$
kink in the $\theta$ variable, is generated by the operator $e^{-\beta
\mu}e^{\pm i2\phi}$. At high $T$ $\phi$ localizes in a minimun of the
cosine and its conjugate field $\theta$ is completely disordered, i.e.
the system looses the superfluid behavior.  The interaction $V(\br)$
between vortices (or charges in the Coulomb-gas
analogy\cite{review_minnaghen,giamarchi_book_1d}) follows from the Gaussian
part of the action \pref{sb}, $V(\br)=\int d^2\bk e^{i\bk\cdot \br}V(\bk)$
where $ V(\bk)=\langle |\phi(\bk)|^2\rangle$, and it is logarithmic since
$V(\bk)=2\pi K/\bk^2$.
%

The physical observables can be easily read out from Eq.\ \pref{sb} and the
free energy $F=-k_BT\ln Z$. For example, the electric current is:
%
%
\be
\lb{js}
\bJ_s(\br,z)=-c\frac{\pd F}{\pd \bA(\br,z)}=-
\frac{2ick_BT}{\Phi_0}\langle \nb\times \hat z\phi(\br)\rangle\delta(z), 
\ee
and it is purely transverse, as expected for vortex excitations. 
The magnetization $\bM=(\bB-\bH)/4\pi$ is defined, as usual, as the
functional derivative of $F$ with respect to $\bB(\br,z)=\nb \times
\bA$. By integrating by part, we can write the last term of Eq.\ \pref{sb} 
as $(2i/\Phi_0) \int d\br dz \ [\bB(\br,z)\cdot \hat z
\phi(\br)]\delta(z)$, so that:
\be
\lb{mag2}
\bM(\br)=-\frac{1}{d}\int dz \frac{\pd F}{\pd \bB(\br,z)}=-\hat z
\frac{2ik_BT}{d\Phi_0}\left\langle\phi(\br)\right\rangle,
\ee
%
which leads to $\bJ_s=c(\nabla \times \bM)$\cite{landau_iv}.  Finally, by
exploiting the fact that $e^{-\beta\mu}e^{\pm i2\phi}$ is the operator
which creates up and down vortices with density $n_\pm$ respectively, we
have a straightforward definition of the average vortex number
$n_F=a^2(\langle n_+\rangle +\langle n_-\rangle)$ and of the excess vortex
number $n=a^2(\langle n_+\rangle -\langle n_-\rangle)$ per unit cell as a
function of $\phi$ as:
\be
\lb{dens}
n_F=2 e^{-\beta\mu} \langle \cos (2\phi)\rangle, \quad
n=2e^{-\beta\mu} \langle \sin (2\phi)\rangle.
\ee
In Eq.\ \pref{mag2} the average value of $\phi$ is computed with the action
\pref{sb}, so that it gives $\bM$ as a function of the magnetic induction
$\bB$. To obtain $\bM$ as a function of the applied field $\bH$ we must use
the Gibbs free energy $\GG=-k_BT\ln Z$, where $Z=\int \DD\phi\DD Ae^{-S}$
and:
$$
S=S_B+\int{d\br dz}\left\{
\frac{(\nb\times\bA)^2}{8\pi k_BT}
-\frac{(\nb\times \bA)\cdot\bH}{4\pi k_BT}\right\}.
$$
$\bH$ satisfies the Maxwell equation $\nb \times \bH=(4\pi/c) \bJ_{ext}$
for a given distribution $\bJ_{ext}$ of external currents.  By integrating
out $\bA$ in the radial gauge $\nb\cdot\bA=0$,
the action reduces to:
\bea
S=\int\frac{d^2\bk}{(2\pi)^2} \frac{k^2+k\L^{-1}}{2\pi K}|\phi(\bk)|^2
-\frac{g}{\pi a^2}\int d\br \cos 2\phi,\nn\\
\lb{sfin}
+\frac{2 i }{\Phi_0} \int d\br\, \phi \,\hat z\cdot \bH^0(\br,z=0)
-\int d\br dz \frac{(\bH^0)^2}{8\pi k_B T},
\eea
where $1/\L=d/2\lambda^2={8\pi^2 Kk_BT}/{\Phi_0^2}$. Here $\bH^0$ is the
magnetic field generated by $\bJ_{ext}$ 
{\em in the vacuum},
i.e. it satisfies the same Maxwell equation as $\bH$, but it is not
constrained to the boundary condition that $\bB=0$
in the SC film. Thus, using the Laplace formula 
$\bH^0(\br)=(1/c)\int d^3\br' [\bJ_{ext}(\br') \times (\br-\br')]/
|\br-\br'|^3$.  The effect of integrating out the $\bB$
field is twofold. First, one introduces an effective screening of the
vortex potential $V(\br)$. Indeed, thanks to the $k\L^{-1}$ term in Eq.\
\pref{sfin}, $V(\br)\sim \log(r/a)$ up to a scale of order $\L$, and then
decays as $\L/r$\cite{review_minnaghen,minnaghen_finitefield}.
Second, one couples directly the dual field $\phi$ to the reference field
$\bH^0$ used in the experiments. We thus expect that in the
Meissner phase $\bM$ includes automatically the demagnetization effects,
i.e.  $-4\pi \bM= \bH^0/(1-\eta)$\cite{landau_iv}, where $\eta$ is the
demagnetization constant which depends only on the sample geometry and
is near to 1 in a film, $\eta\sim 1-d/R$\cite{landau_iv,fetter_films}, where
$R$ is the transverse film dimension.


The model \pref{sfin} and the constitutive equations \pref{js}-\pref{dens}
establish a clear and general theoretical framework to address the physics
of 2D SC films in a magnetic field. To illustrate their usefulness we solve
them by using a variational approximation. The idea is
to replace the cosine interaction in Eq.\ \pref{sfin} with a mass term
$\D^2\phi^2$, where $\D$ is determined self-consistently by minimizing the
variational energy $\GG_{var}=\GG_0+T\langle S-S_0\rangle$, $S_0$ being the
trial action.  At $\bH^0=0$ a finite $\D$ appears above $T_{BKT}$, which
signals the localization of $\phi$ in a minimum of the cosine, and cut-off
at a scale $1/\D$ the logarithmic vortex potential $V(\br)$. This allows for
the proliferation of free-vortex excitations.  We then consider the case of
a perpendicular field $\bH=H\hat z$ (in the following we drop the
superscript $0$) slowly varying over the film. To account for it we
introduce in the trial action an additional variational parameter $\bar H$,
coupled linearly to $\int d\br\phi$ in analogy with Eq.\ \pref{sfin}, so
that only the $\phi(k)$ component at the minimum $k$ value $k_{min}\simeq
1/R$ couples to $H$.  A finite system size $R$ is needed to have finite
demagnetization in the Meissner phase, but its role at large fields (and in
general above $T_{BKT}$) is negligible. The trial action is:
%
\be
\lb{trial2}
S_0=\frac{1}{2\O}\sum_\bk \left[G^{-1}(\bk)
\phi(\bk)\phi(-\bk)\right]+\frac{2i}{\Phi_0}\bar H\phi(\bk_{min}),
\ee
where $\O\sim R^2$ is the film area and $G^{-1}(\bk)=(
k^2+k\L^{-1}+(\D/a)^2)/{\pi K}$. According to Eq.\ \pref{mag2} the
magnetization is related to $\D$ and $\bar H$ as
\be
\lb{magh}
M=-\frac{k_B T}{d\phi_0} \frac{4\pi K}{\D^2+\D_R^2}
\frac{\bar H a^2}{\Phi_0}\equiv
-\frac{k_B T}{d\Phi_0} \Mt
\ee
where $\Mt$ is the dimensionless magnetization and 
$(\D_R/a)^2=1/R^2+1/R\L$ is the intrinsic (i.e. $T$ and $H$
independent) cut-off. By minimizing $\GG_{var}$ with respect to
$(\D,\bar H)$ we derive the two self-consistent equations:
\bea
\lb{sc1h}
& &4Kg (\D+\D_\L)^K\cosh(\Mt)=
\D^2\\
\lb{sc2h}
& &\D^2\tanh(\Mt)=
n_H(4\pi K)-\Mt\D_R^2,
\eea
where $\D_\L=\D_R^2R/a$ and $n_H=Ha^2/\Phi_0$ is the flux per
unit cell.
Finally, Eq.\ \pref{dens} leads to:
%
\be
\lb{defnfh}
n_F={\D^2}/{4\pi K}, \quad n=n_H-\Mt {\D_R^2}/{4\pi K}.
\ee
We note that using Eq.\ \pref{defnfh} the two Eqs.\ \pref{sc1h}-\pref{sc2h} can
be related to similar expressions derived in Ref.\
\cite{minnaghen_finitefield,doniach_films}. Nonetheless, a clear connection
to the magnetization and to the role of $\bH$ vs $\bB$ was  lacking in
these papers. 
\begin{figure}[htb]
\includegraphics[scale=0.35,angle=-90]{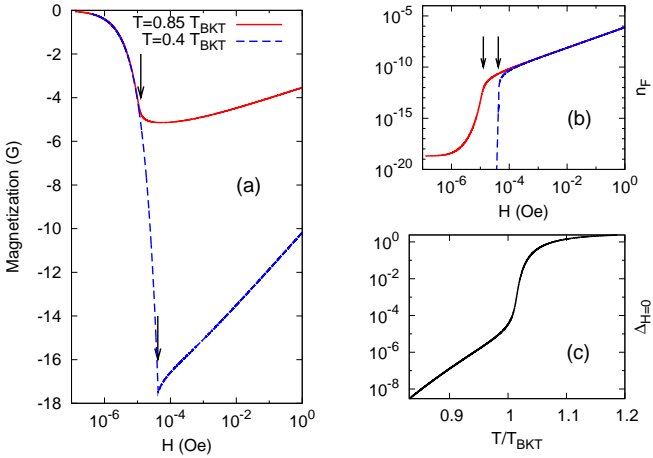}
\caption{(Color online) (a)-(b) $M(H)$ and $n_F(H)$ below $T_{BKT}$ from
  the numerical solution of Eqs.\ \pref{sc1h}-\pref{sc2h}. The arrows
  indicate $H_{c1}$ according to Eq.\ \pref{hc1}. At $T=0.4 T_{BKT}$ $M(H)$
  shows a sharp kink at $H_{c1}$, where $n_F$ drops abruptly to small
  values. At higher $T$ these features are partly smoothened out by thermal
  smearing. (c) Temperature dependence of the mass term $\D_{H=0}$. Notice
  that $\D_{H=0}\sim \D_\L$ at $T_{BKT}$, where $\L=6.9 \times 10^5$ \AA,
  $\D_R\sim 4\times 10^{-6}$ and $\D_\L=1.9\times 10^{-5}$.}
\label{magnet}
\end{figure}
As a prototype of 2D system we consider a single layer of underdoped
Bi2212, with $J(T)=J_0(1-T/T_{MF})$ to mimic the bare $T$ dependence due to
quasiparticles, $J_0=180$ K and $T_{MF}=120$ K, which gives $T_{BKT}=84$
K. For $d=15$ \AA \ as the typical interlayer distance the magnetization
\pref{magh} is given in units of $k_B T/d\Phi_0=(4.4\times 10^{-3} T)$ G
(or $(4.4 \times T)$ A/m in the notation of Ref.\
\cite{li_magn}). Moreover, we use $R\sim 10^6 a$ as the typical sample
size, with $a=40$ \AA, and choose $\mu=1.5 \mu_{XY}$.
With this choice of parameters one has always $R\gg \L$ , so that $\D_\L\sim
1/\L\gg \D_R\sim 1/\sqrt{R\L}$ up to $T_{BKT}$. 

At $H=0$ Eq.s \pref{sc1h}-\pref{sc2h} are satisfied for $\Mt=0$ and $\D$
solution of the equation $ 4Kg(\D_{H=0}+\D_\L)^K=\D_{H=0}^2$. At $\D_\L=0$
the solution $\D_{H=0}^2=(4Kg)^{2/(2-K)}$ is finite only at $K<2$, which
identifies the BKT transition at $K=2$ (i.e. $T_{BKT}=\pi
J(T_{BKT})/2$). When a finite cut-off $\D_\L$ is introduced $\D_{H=0}$
approaches $\D_\L$ at $T_{BKT}$, and vanishes as
$\D_{H=0}=\sqrt{4Kg}\D_\L^{K/2}$ as $T\ra0$, giving $\D\ll \D_\L,\D_R$
already at $T\lesssim 0.9 T_{BKT}$, see Fig.\ 1c. Observe that at $H=0$,
where the same number of $\pm$ vortices are thermally-induced, $n=0$, and one
can
parametrize $n_F$ in Eq.\ \pref{defnfh} via the vortex correlation length
$\xi$ as $1/\xi^2\equiv n_F/a^2=\D_{H=0}^2/4\pi K a^2$.

At $H\neq 0$ a finite $\Mt$ appears, which modifies also the $\D$ value. In
general, at low field $\D$ keeps the zero-field value $\D(H)\approx
\D_{H=0}$ and $\Mt$ grows linearly with $H$. By further increasing $H$,
$\D$ grows with respect to $\D_{H=0}$ and $\Mt$ enters a non-linear
regime.  The slope of $M$ vs $H$, the absolute value of $\Mt$ and the
crossover field differ substantially above and below $T_{BKT}$.  Let us
first analyze the case $T<T_{BKT}$, i.e. $K>2$. For small $\Mt$ one has
$\tanh(\Mt)\approx \Mt$ and $\cosh(\Mt)\approx 1$, so that we obtain
$\Mt=n_H(4\pi K)/(\D^2+\D_R^2)$ and $\D(H)\approx \D_{H=0}$ from Eq.\
\pref{sc2h} and Eq.\ \pref{sc1h}, respectively.  Since $\D_{H=0}\ll
(\D_R,\D_\L)$ as $T\lesssim 0.9 T_{BKT}$, and $\D_R^2\sim a^2/R\L$, we obtain
(using $8\pi^2K k_BT/\Phi_0^2=1/\L$):
\be
\lb{dem}
M=-\frac{1}{4\pi}\frac{2R}{d} H, \quad
n=n_H\frac{\D_{H=0}^2}{\D_R^2}\approx 0,
\ee
where we recognize flux expulsion ($n\equiv Ba^2/\Phi_0\approx 0$) and the
Meissner effect ($-4\pi M= H/(1-\eta)$) in the presence of a large
demagnetization factor $\eta\sim 1-d/R$ as expected in a thin
film\cite{landau_iv,fetter_films}.  At large field instead
$\cosh(\Mt)\approx e^{\Mt}/2$ and $\tanh(\Mt)\approx 1$. We then obtain
that $\Mt\approx \log(2\D^{2-K}/p)$ from Eq.\ \pref{sc1h}, and using
$\D^2\approx 4\pi K n_H$ from Eq.\ \pref{sc2h} we get:
\be
\lb{mlog}
M=-\frac{k_BT}{d\Phi_0}\left[A(T)+\left(\frac{T_{BKT}}{T}-1\right)
\log\left(\frac{\Phi_0}{a^2 H}\right)\right],
\ee
with $A(T)=\mu/k_BT-(K/2)\log(4\pi K)$.  The linear regime \pref{dem}
survives up to a field $H_l^b$ that can be determined by the numerical
solution of Eqs.\ \pref{sc1h}-\pref{sc2h}. As it is shown in Fig.\ 2b,
$H_l^b$ is very low ($\sim 10^{-6}$ G) but finite at $T_{KT}$.  For this
reason,  the field-independence of $M$ at criticality implied by Eq.\
\pref{mlog} is only valid above $H_l^b$, below which $M\propto -H$, as
expected.  This low-field crossing to a linear behavior is missing in Ref.\
\cite{sondhi_kt} where $M$ is calculated as a function of $B$. However, at
large fields where $B\approx H$ the dependence of $M(B)$ on $\log(B)$
derived there coincides with Eq.\ \pref{mlog}, apart from an additional $B$
dependence of $M$ at criticality that cannot be checked with the present
variational calculation.
Finally, we notice that at $T$ well below $T_{BKT}$ an
estimate of $H_l^b$ can be obtained analytically by matching the high-field
and low-field solutions for $\Mt$ at $\D\approx \D_\L$:
\be
\lb{hc1}
H_{c1}=\frac{\Phi_0}{4\pi }
\frac{(\D_R/a)^2}{K}
\left[ (2-K)\log(2\D_\L)-
\log(8Kg)\right],
\ee
which reduces for $T\ra 0$ to the standard definition of first
critical field in a SC film, $H_{c1}(T\ra 0)=(\Phi_0/4\pi \L R)
\log(\L/2a)+4\pi\mu/\Phi_0 R$\cite{fetter_films}. Indeed, as we can see in
Fig.\ 1a, at low $T$ the magnetization displays a sharp kink at $H_{c1}$
and increases just above it, as indeed expected at the threshold of flux
penetration (see also $n_F$ in Fig.\ 1b). However, at higher temperatures
such a kink in $M$ disappears due to thermal smearing and the minimum of
$M$ is located at a field higher than $H_{c1}$. 

%
%
\begin{figure}[htb]
\includegraphics[scale=0.35,angle=-90]{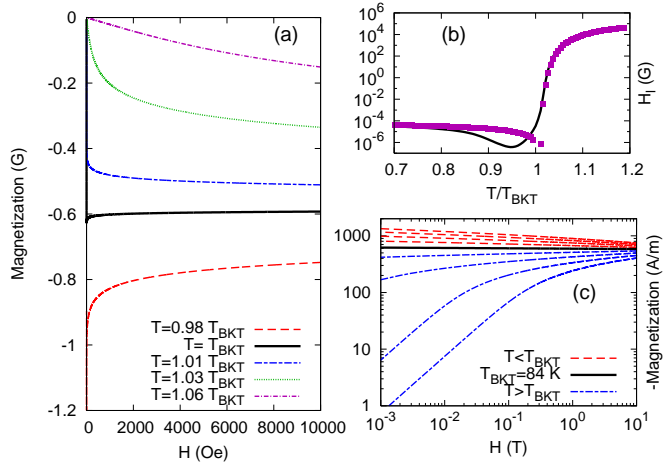}
\caption{(Color online) (a) $M(H)$ above and below $T_{BKT}$ from the
numerical solution of Eqs.\ \pref{sc1h}-\pref{sc2h}. 
(b) Solid line: the threshold field
$H_l^{a,b}$ as a function of $T$. The points
show the analytical estimates \pref{hc1}-\pref{hlin}, which agree with the
numerical result except in a small range near $T_{BKT}$.  (c) $M(H)$ in
logarithmic scale (curves are spaced by 1K).}
\label{sol_h=0}
\end{figure}

At $T>T_{BKT}$, i.e. $K<2$, $\Mt$ shows again a crossover from a linear to
non-linear behavior at a field $H_l^a$. To estimate $H_l^a$ we can expand
the hyperbolic functions in Eq.s \pref{sc1h}-\pref{sc2h} around $\Mt=0$.
For $T$ sufficiently above $T_{BKT}$ so that $\D^2_{H=0}\simeq
(4Kg)^{2/(2-K)}\gg (\D_\L,\D_R)$ we obtain the approximate solutions
$\Mt=(4\pi K n_H)/\D^2$ and $
\D^2=\D_{H=0}^2\left[1+1/({2-K})(H\xi/\Phi_0)^2\right]$, where $\xi$ is the
zero-field correlation length defined above. When the second term in the
square brackets is $\ll 1$ the deviations of $\D^2$ with respect to
$\D_{H=0}^2$ are negligible, so that
\be
\lb{hlin}
M=-\frac{k_B T}{d\Phi_0^2}\xi^2 H, \, H\lesssim 
H_l^a=0.1 \frac{\Phi_0}{\xi^2}\sqrt{\frac{T-T_{BKT}}{T}}
\ee
At $T$ sufficiently close to $T_{BKT}$ screening effects cut-off both $\D$
(i.e. $\xi$) and $H_l^a$, so that the estimate \pref{hlin} is no more
valid, $H_l^a$ attains a finite value and merges with the field $H_l^b$
discussed above, see Fig.\ 2b. As it was
known\cite{halperin_ktfilms,sondhi_kt} the functional dependence of the
low-field magnetization $M$ on the BKT correlation length $\xi$ in Eq.\
\pref{hlin} is the same as in the GL theory\cite{koshelev_gaussian}.
However, the critical region $H<H_l^a$ where 
such a dependence is valid turns out to be remarkably
smaller than in the standard GL theory\cite{koshelev_gaussian}, because
$\xi$ diverges much faster than in the GL case as $T\ra T_{BKT}$.

In conclusion, we proposed a new theoretical framework to investigate the
KT physics of 2D superconductors in a finite magnetic field, as given by
the modified sine-Gordon model \pref{sfin} and the definitions
\pref{js}-\pref{dens} of the physical quantities as a function of the
applied magnetic field $\bH$ (instead of $\bB$). As we showed within a
variational analysis of the model \pref{sfin}, we obtain a clear
description of the Meissner phase below $T_{BKT}$, and an estimate of the
threshold field $H_l^{a,b}$ for the appearance of non-linear effects. Above
$T_{BKT}$ the shrinking of the linear regime with respect to standard GL
fluctuations is a typical signature of the faster divergence of $\xi$
within the BKT theory.  These results can shed new light on the
physics of vortices in cuprates.
Indeed, taking into account that in layered superconductors the intrinsic
cut-off $R_J$ is provided by the interlayer coupling $J_\perp$ instead of
$\L$, $R_J\sim a\sqrt{J/J_\perp}$, our 2D calculations can be applied to
these systems in all the $(T,H)$ range where $\D\gg 1/R_J$ (so that for
example large demagnetization effects are not expected in layered
systems). Thus, the persistence of a non-linear magnetization
up to $H\sim 0.01$ T in a wide range of temperatures above $T_{BKT}$ found
experimentally in Ref.\ \cite{li_magn} can be a signature of the rapid
decreasing of $H_l^a$ as $T\ra T_{BKT}$, which does not contradict but
eventually support the KT nature of the SC fluctuations in these
systems. Moreover, since $\xi$ increases as $\mu$ increases, the extremely
low values of $H_l^a$ measured in Ref.\ \cite{li_magn} suggest a value of
$\mu$ larger than $\mu_{XY}$, in agreement with the result of Ref.\
\cite{benfatto_BKT_in_cuprates} based on the analysis of the superfluid
density, and call for a deeper investigation of the normal phase existing
in the vortex cores.

\vspace{-0.5cm}

%

%
%


\end{document}